\documentclass[pra,twocolumn,showpacs,preprintnumbers,superscriptaddress]
{revtex4}

\usepackage{times}
\usepackage{bm}
\usepackage{float}
\usepackage{graphicx}
\usepackage{amsbsy}
\usepackage{amsmath}
\usepackage{amsfonts}
\usepackage{amsthm}
\usepackage{xcolor}

\begin{document}
	\theoremstyle{plain}
	\newtheorem{theorem}{Theorem}
	\newtheorem{lemma}[theorem]{Lemma}
	\newtheorem{corollary}[theorem]{Corollary}
	\newtheorem{proposition}[theorem]{Proposition}
	\newtheorem{conjecture}[theorem]{Conjecture}
	
	\theoremstyle{definition}
	\newtheorem{definition}[theorem]{Definition}
	
	\theoremstyle{remark}
	\newtheorem*{remark}{Remark}
	\newtheorem{example}{Example}
	\title{Physical realization of realignment criteria using structural physical approximation}
	\author{Shruti Aggarwal, Anu Kumari, Satyabrata Adhikari}
	\email{shruti_phd2k19@dtu.ac.in, mkumari_phd2k18@dtu.ac.in, satyabrata@dtu.ac.in} \affiliation{Delhi Technological
		University, Delhi-110042, Delhi, India}
	
	\begin{abstract}
		Entanglement detection is an important problem in quantum information theory because quantum entanglement is a key resource in quantum information processing. Realignment criteria is a powerful tool for detection of entangled states in bipartite and multipartite quantum system. It works well not only for negative partial transpose entangled states (NPTES) but also for positive partial transpose entangled states (PPTES). Since the matrix corresponding to realignment map is indefinite so the experimental implementation of the map is an obscure task. In this work, firstly, we have approximated the realignment map to a positive map using the method of structural physical approximation (SPA) and then we have shown that the structural physical approximation of realignment map (SPA-R) is completely positive. Positivity of the constructed map is characterized using moments which can be physically measured. Next, we develop a separability criterion based on our SPA-R map in the form of an inequality and have shown that the developed criterion not only detect NPTES but also PPTES. Further we have shown that for a special class of states called Schmidt symmetric states, the SPA-R separability criteria reduces to the original form of realignment criteria. We have provided some examples to support the results obtained. Moreover, we have analysed the error that may occur because of approximating the realignment map. 
	\end{abstract}
	\pacs{03.67.Hk, 03.67.-a} \maketitle

	\section{Introduction}
	Entanglement \cite{horodeckirev} is a key ingredient in quantum physics and the future of quantum technologies. It has advantages in various quantum information processing tasks such as quantum communication\cite{bennett,bennett2,ekert}, quantum computation\cite{nielsen}, remote state preparation\cite{pati}, quantum simulation\cite{lioyd} and thus, detection of entanglement is an important problem in quantum information theory. Detection of entanglement is also important because even if an experiment is carried out to generate entangled state in bipartite or multipartite quantum system, the generated state may not be entangled due to the presence of noise in the environment and it is quite difficult to check whether the generated state is entangled or not. Despite many efforts, a complete solution for the separability problem is still not known.
		Positive maps are strong detectors of entanglement. However, not every positive map can be regarded as physical, for example, in case of describing a quantum channel or the reduced dynamics of an open system, a stronger positivity condition is required \cite{kraus}. Completely positive maps play an important role in quantum information theory, since a positive map is physical whenever it is completely positive. Completely positive maps were introduced by Stinespring in the study of dilation problems for operators \cite{stein}. Let $B(\mathcal{H}_A)$ and $B(\mathcal{H}_B)$ denote the set of bounded operators on the Hilbert spaces $\mathcal{H}_A$ and $\mathcal{H}_B$, respectively. If the Hilbert space $\mathcal{H}= \mathcal{H}_A  \otimes\mathcal{H}_B$ has dimension $k$, we identify $B(\mathcal{H})$ with $M_k(\mathbb{C})$, the space of $k \times k$ matrices in $\mathbb{C}$. A linear map $\Phi: B(\mathcal{H}_A) \longrightarrow B(\mathcal{H}_B)$ is positive if $\Phi(\rho)$ is positive for each positive $\rho \in B(\mathcal{H}_A) $. The map $\Phi$ is completely positive, if for each positive integer $k$, the map $I_k \otimes \Phi : M_k (B(\mathcal{H}_A)) \longrightarrow M_k (B(\mathcal{H}_B))$ is positive.
	In quantum information theory, completely positive maps are important because they are used to characterize quantum operations \cite{nielsen}. Choi described the operator sum representation of completely positive maps in \cite{choi}. In \cite{poon}, necessary and sufficient conditions for the existence of completely positive maps are given. A physical way by which positive maps can be approximated by completely positive maps is called structural physical approximation (SPA) \cite{horodecki3,jaro, augusiak11,hakye,augusiak14,jbae}. The idea is to mix a positive map $\Phi$ with maximally mixed state, making the mixture $\tilde{\Phi}$ completely positive \cite{horodecki3}. The resulting map can then be physically realized in a laboratory and its action characterizes entanglement of the states detected by $\Phi$. In addition, the resulting map keeps the structure of the output of the non physical map $\Phi$ since the direction of generalized Bloch vector of the output state remaims same as the output state of the original nonphysical map, only the length of the vector is rescaled by some factor \cite{horo2003}. The SPA to the map $\Phi$ in $d\otimes d$ dimensional space is given by 
\begin{equation}
	\tilde{\Phi}(\rho) = \frac{p^*}{d^2} I_{d^2} + (1-p^*)\Phi(\rho)
\end{equation}
where $I_{d^2}$ denotes the identity matrix of order $d^2$ and $p^*$ is the minimum value of the probability $p$ for which the approximated map $\tilde{\Phi}$ is completely positive \cite{adh}.\\
Although, there exist various methods in literature for detection of entangled states, the first solution to this problem is connected to the theory of positive maps. It was proposed by Peres in the form of partial transposition (PT) criteria  \cite{peres}. Later, Horodecki proved that this criteria is necessary and sufficient for $2\otimes 2$ and $2\otimes 3$ dimensional quantum system \cite{horodecki2}. Although this criteria is one of the most important and widely used criteria but it suffers from serious drawbacks. One of the major drawback  is that it is based on the negative eigenvalues of the partially transposed matrix and thus used to detect negative partial transpose entangled states (NPTES) only. Another drawback is that the partial transposition map is positive but not a completely positive map and hence, may not be implemented in an experiment. In order to make it experimentally implementable, partial transposition map have been approximated to a completely positive map using the method of SPA \cite{horodecki3}. A lot of work have been done on the structural physical approximation of partial transposition (SPA-PT) \cite{horodecki3,adh,hlim1, hlim3, kumari1}. The SPA-PT have been used to detect and quantify entanglement \cite{horodecki3,adhikari1} but till now it can only be used to detect and quantify NPTES only. \\
As the PPT criterion fails to detect bound entanglement in higher dimensions, certain other criterion have been proposed in the literature which can detect some positive partial transpose entangled states (PPTES). These include the realignment criteria or computable cross-norm criteria (CCNR) \cite{chen, rudolph5}, range criterion \cite{terhal}, covariance matrix criterion \cite{hyllus}.
 Moreover, it has been shown that the PPT criterion and the CCNR criterion are equivalent under permutations of the density matrix’s indices \cite{rudolph2004}. The generalization of CCNR criterion were investigated in \cite{chen2}. The symmetric function of Schmidt coefficients have been used to improve the CCNR criterion in \cite{lupo}. Separability criteria based on the realignment of density matrices and reduced density matrices have been proposed in \cite{shen}. In \cite{shruti1,shruti2}, witness operators using the realignment map is constructed which efficiently detect and quantify PPT entangled states.
 In \cite{XQi}, rank of realigned matrix is used to obtain a necessary and sufficient product criteria for quantum states.
 Recently, methods for detecting bipartite entanglement based on estimating moments of the realignment matrix have been proposed \cite{tzhang,shruti4}. Realignment criteria is a powerful criteria in the sense that it may be used to detect NPTES as well as PPTES. PPTES, also known as bound entangled states, which are weak entangled states that cannot be distilled by performing local operations and classical communications (LOCC). Although it is one of the best for the detection of PPTES but the problem with this criteria is that it may not be used to detect entanglement practically because the realignment map corresponds to a non-positive map and it is known that the non-positive maps are not experimentally implementable. Also, it is known that completely positive maps may be realized in an experiment \cite{korbicz}. The defect that the realignment map may not be realized in an experiment may be overcome by approximating the non-positive realignment map to a completely positive map. Our work is significant because although there has been considerable progress in entanglement detection using the SPA of partial transposition map but the idea of SPA of realignment operation is still unexplored.  \\
	In this work we approximate the non-positive realignment map to a completely positive map. To achieve this goal, we first approximate the non-positive realignment map with a positive map and then we show that the obtained positive map is also completely positive. We estimate the eigenvalues of the realignment matrix using moments which may be used physically in an experiment \cite{brun,tanaka,sougato,imai}. Further, we formulate a separability criterion that we call SPA-R criteria, using our approximated map that not only detects NPT entangled states but also PPT entangled states.  Next, we have shown that the SPA-R criteria reduces to the original formulation of realignment criteria for a class of states called Schmidt-symmetric states.  Moreover we discuss the accuracy of our approximated realignment (SPA-R) map by calculating the error of the approximation in trace norm. We also introduce an error inequality which holds for all separable states.
	
	This paper is organized as follows: In Sec. II, we will revisit realignment criteria and review some preliminary results that we will use in further sections. In Sec. III, we approximate the non-positive realignment map to a positive map and further, we will show that the approximated positive map is completely positive. In Sec. IV, we develop our separability criteria called SPA-R criteria based on approximated realignment map. Furthermore, we show that the SPA-R criteria and the original form of realignment criteria will become same for Schmidt symmetric states. In Sec. V, we investigate the error generated due to the approximation procedure. In Sec. VI, we illustrate some examples to support the results obtained in this work. In Sec. VII, we discuss the efficiency of SPA-R criteria. Finally, we conclude in Sec. VIII.
	\section{Preliminaries}
	In this section, we stated the realignment criteria and some results which are discussed in the literature. We will use these results in the subsequent section to obtain the modified form of realignment criteria that may be realizable in the experiment.\\
	\subsection{Realignment Criteria}
	First, let us recall the definition of the realignment operation. For any $m \times m$ block matrix $X$ with each block $X_{ij}$ of size $n \times n$, $i, j = 1, \; . \;.\;., m$, the realigned matrix $R(X)$ is defined by
	\begin{eqnarray}
		R(X)= [vec(X_{11}), . . ., vec(X_{m1}), . . ., vec(X_{1m}), . . ., vec(X_{mm})]^t \nonumber\\
	\end{eqnarray}
where for any $n \times n$ matrix $X_{ij}$ with entries $x_{ij}$, $vec(X_{ij})$ is defined as
\begin{eqnarray}
	vec(X_{ij})= [x_{11}, . . ., x_{n1}, x_{12} . . ., x_{n2}, . . ., x_{1n}, . . ., x_{nn}]^t
\end{eqnarray}
\\
	Let us consider a bipartite quantum system described by a density operator $\rho$ in $H_A^{d_1}\otimes H_B^{d_2}$ dimensional quantum system. The density operator $\rho$ may be expressed as
	\begin{eqnarray}
	\rho=\sum_{i,j,k,l}p_{ij,kl}|ij\rangle \langle kl|
	\end{eqnarray}
	where $d_1$ and $d_2$ are the dimensions of the Hilbert spaces $H_A$ and $H_B$ respectively. After applying realignment operation on $\rho$, the realigned matrix $R(\rho)$ may be expressed as
	\begin{eqnarray}
	R(\rho)=\sum_{i,j,k,l}p_{ij,kl}|ik\rangle \langle jl|
	\end{eqnarray}  
	Then realignment criteria may be stated as: If $\rho$ represent a separable state then $||R(\rho)||_1 \leq 1$. Here $||.||_1$ denotes the trace norm and it may be defined as $||T||_1=Tr(\sqrt{TT^{\dagger}})$ \cite{chen}.\\

\subsection{A few well-known results}
In this section, we mention about a few important results that may require in the following section. To proceed with, we employ a useful theorem by Weyl \cite{weyl}, that connects the eigenvalues of the sum of Hermitian matrices to those of the individual matrices. We use this theorem to prove the positivity of our approxiamted map. For convenience, Weyl's theorem can be stated as follows:\\
\textbf{Result 1:} (Weyl's Inequality \cite{weyl}) Let $A,B\in M_n$ be two Hermitian matrices and let $\{\lambda_i[A]\}_{i=1}^n$, $\{\lambda_i[B]\}_{i=1}^n$ and $\{\lambda_i[A+B]\}_{i=1}^n$ be the eigenvalues of $A,B$ and $A+B$ respectively, arranged in ascending order, i.e., $\lambda_1 \leq \lambda_2 \leq. . . \leq \lambda_n$. Then 
\begin{eqnarray}
	(i)~~\lambda_k[A+B]\leq \lambda_{k+j}[A]+\lambda_{n-j}[B],~~j=0,...,n-k 
		\end{eqnarray}
\begin{eqnarray}
	(ii)~~\lambda_{k-j+1}[A]+\lambda_j[B]\leq \lambda_k[A+B],~~j=1,...,k 
		\label{weyl}
\end{eqnarray}
It may not be an easy task to directly compute the eigenvalues of a matrix, thus bounds for eigenvalues are of great importance. Bounds for eigenvalues using traces have been studied in \cite{wolko}. Further, the bound of the eigenvalues expressed in terms of moments may be useful for the experimentalist to estimate eigenvalues in the laboratory. We now state the result \cite{wolko} given below that determine a lower bound for the minimum eigenvalue of a matrix in terms of first and second order moments of the matrix. We will use the following result 2 in the subsequent section to prove the positivity of our approximated map.\\  
\textbf{Result 2 \cite{wolko}}: Let $A \in M_n(\mathbb{C})$ be any matrix with real eigenvalues and $\lambda_{min}^{lb}[A]$ denotes the lower bound of the minimum eigenvalue of $A$. Then
\begin{eqnarray}
	\lambda_{min}^{lb}[A] \leq \lambda_{min}[A] \
\end{eqnarray}
where the lower bound is given by 
\begin{eqnarray}
	\lambda_{min}^{lb}[A] =	\frac{Tr[A]}{n} - \sqrt{(n-1)\left(\frac{Tr[A^2]}{n}- (\frac{Tr[A]}{n})^2\right)}\nonumber\\ \label{lb}
\end{eqnarray}
The useful conditions for the existence of completely positive maps have been studied in \cite{poon}. We have exploited the conditions to prove the completely positivity of the introduced approximated map. The conditions may be expressed as the Result 3 below.\\
\textbf{Result 3 \cite{poon}:}  Consider a map $\Phi: M_n(\mathbb{C}) \longrightarrow M_m(\mathbb{C})$. Let $A \in M_n$ and $B \in M_m$ be Hermitian matrices such that $\Phi(A)=B$. Then the map $\Phi$ is completely positive iff there exist non-negative real numbers $\gamma_1$ and $\gamma_2$ such that the following conditions hold:
\begin{eqnarray}
	\lambda_{min}[B]&\geq& \gamma_1 \lambda_{min}[A]\\
	\lambda_{max}[B]&\leq& \gamma_2 \lambda_{max}[A]
\end{eqnarray}

\section{Structural Physical Approximation of Realignment Map: Positivity and Completely positivity}
\noindent In this section, we  employ the method of structural physical approximation to approximate the realignment map. To proceed toward our aim, let us first recall the depolarizing map which may be defined in the following way:\\
A map $\Phi_d: M_n \longrightarrow M_n$ is said to be depolarizing if 
\begin{eqnarray}
\Phi(A)= \frac{Tr[A]}{n} I_{n}
\label{dep}
\end{eqnarray}
In the method of structural physical approximation, we mix an appropriate proportion of realignment map with a depolarizing map in such a way that the resulting map will be positive. This may happen because the lowest negative eigenvalues generated by the realignment map can be offset by the  eigenvalues of the maximally mixed state generated by the depolarizing map.\\
Consider any quantum state $\rho$ in $d \otimes d$ dimensional system $\mathcal{D} \subset \mathcal{H}_A \otimes \mathcal{H}_B$ such that $\mathcal{D}$ contains the states $\rho$ whose realignment matrix $R(\rho)$ have real eigenvalues and positive trace. The structural physical approximation of realignment map may be defined as $\widetilde{R}: M_{d^2} (\mathbb{C}) \longrightarrow M_{d^2}(\mathbb{C})$ such that
\begin{eqnarray}
\widetilde{R}(\rho)=\frac{p}{d^2}I_{d\otimes d}+\frac{(1-p)}{Tr[R(\rho)]}{R(\rho)},~~0 \leq p \leq 1
\label{sparealign}
\end{eqnarray}

\subsection{Positivity of structural physical approximation of realignment map}
 It is known that $R(\rho)$ forms an indefinite matrix, its eigenvalues may be negative or positive. Let us first consider the case when all the eigenvalues of $R(\rho)$ are non negative. So, by the definition of $\widetilde{R}$ given in (\ref{sparealign}), $\widetilde{R}(\rho)$ is positive for all $ p \in [0,1]$ and hence $\widetilde{R}$ defines a positive map. On the other hand if $R(\rho)$ has negative eigenvalues, then $\widetilde{R} (\rho)$ may be positive under some conditions.
But since the realignment operation is not physically realizable, it is not feasible to compute the eigenvalues of $R(\rho)$. To overcome this challenge, we find the range of $p$ in terms of $\lambda_{min}^{lb}[R(\rho)]$ defined in (\ref{lb}), which can be expressed in terms of $Tr[R(\rho)]$ and $Tr[(R(\rho))^2]$. The first and second moments of $R(\rho)$ may be measured experimentally \cite{sougato}. Now the problem is: how to determine the sign of the real eigenvalues of $R(\rho)$ experimentally without directly computing its eigenvalues? The method we develop here to tackle this problem is described below in detail.


\subsubsection{Method for determining the sign of real eigenvalues of $R(\rho)$}
Let $\rho \in \mathcal{D}$ be a $d \otimes d$ dimensional state such that $R(\rho)$ has real eigenvalues $\lambda_1, \lambda_2, . . ., \lambda_{d^2}$. The characteristic polynomial of $R(\rho)$ is given as 
\begin{eqnarray}
	f(x) = \prod_{i=1}^{d^2} (x - \lambda_i) = \sum_{k=0}^{d^2} (-1)^k a_k x^{d^2-k}
\end{eqnarray} 
where $a_0 = 1$ and $\{a_k\}_{k=1}^{d^2}$ are the functions of eigenvalues of $R(\rho)$.\\
Let us now consider the polynomial $f(-x)$, which
effectively replaces the positive eigenvalues of $R(\rho)$ by negative ones and vice versa. 
For a polynomial with real roots, Descartes’ rule of
sign states that the number of positive roots is given by the number of sign changes between consecutive elements in the ordered list of its nonzero coefficients \cite{descartes}. The matrix $R(\rho)$ is positive semi-definite iff the number of sign change in the ordered list of non-zero coefficients of $f(x)$ is equal to the degree of the polynomial $f(x)$. These non-zero coefficients can be determined in terms of moments of the matrix $R(\rho)$. The coefficients $a_i$'s are related to the moments of $R(\rho)$ by the recursive formula \cite{newton}
\begin{eqnarray}
	a_k = \frac{1}{k} \sum_{i=0}^k (-1)^{i-1} a_{k-i}  m_i (R(\rho)) \label{rec}
\end{eqnarray}
where $ m_i (R(\rho)) = Tr[(R(\rho))^i]$ denotes the $ith$ order moment of the matrix $R(\rho)$. For convenience, we write $ m_i (R(\rho))$ as $m_i$. The ith order moment can be explicitly expressed as 
\begin{eqnarray}
	m_i  = (-1)^{i-1} i a_i + \sum_{k=1}^{i-1} (-1)^{i-1+k} a_{i-k} m_k
\end{eqnarray}

Using (\ref{rec}), we get
\begin{eqnarray}
	a_1 &=& m_1\\
	a_2 &=& \frac{1}{2} (m_1^2 - m_2) \\
	a_3 &=& \frac{1}{6} ( m_1^3 - 3 m_1 m_2 + 2m_3)
\end{eqnarray} and so on.\\
Therefore, the matrix $R(\rho)$ is positive semi-definite iff $a_i \geq 0$
for all $i = 1, ..., d^2$. \\

\subsubsection{Positivity of $\widetilde{R}(\rho)$:}
In this section, we derive the condition for which the approximated map $\tilde{R}(\rho)$ will be positive when (i) $R(\rho)$ is positive; and when (ii) $R(\rho)$ is indefinite. The obtained conditions are stated in the following theorem.\\
\textbf{Theorem-1} Let $\rho$ be a $d \otimes d$ dimensional state such that its realignment matrix  $R(\rho)$ has real eigenvalues. The structural physical approximation of realignment map $\widetilde{R}(\rho)$ is a positive operator for $p \in [l, 1]$, where $l$ is given by
\begin{eqnarray}
l= \left\{
\begin{array}{lrr}
0 & when & \lambda_{min} [R(\rho)] \geq 0 \\
\frac{d^2k}{Tr[R(\rho)]+d^2k} \leq p \leq 1 & when & \lambda_{min} [R(\rho)] < 0
\end{array}\right\}
\label{thm1}
\end{eqnarray}
where $k=max[0,-\lambda_{min}^{lb}[R(\rho)]]$
and $\lambda_{min}^{lb}[R(\rho)])$ denotes the lower bound of the minimum eigenvalue of $R(\rho)$ defined in (\ref{lb}).\\
\textbf{Proof:} Recalling the definition (\ref{sparealign}) of the SPA of the realignment map, the minimum eigenvalue of $\widetilde{R}(\rho)$ is given by
\begin{eqnarray}
\lambda_{min}[\widetilde{R}(\rho)]=\lambda_{min}[\frac{p}{d^2}I_{d\otimes d}+\frac{(1-p)}{Tr[R(\rho)]}{R(\rho)}]
\label{minr}
\end{eqnarray}
where $\lambda_{min}(.)$ denote the minimum eigenvalue of $[.]$. Using Weyl's inequality given in (\ref{weyl}) on RHS of (\ref{minr}), it reduces to
\begin{eqnarray}
\lambda_{min}[\widetilde{R}(\rho)]&\geq& \lambda_{min}[\frac{p}{d^2}I_{d\otimes d}]+\lambda_{min}[\frac{(1-p)}{Tr[R(\rho)]}{R(\rho)}]\nonumber\\&=& \frac{p}{d^2}+\frac{(1-p)}{Tr[R(\rho)]}\lambda_{min}{[R(\rho )]}
\label{lambmin}
\end{eqnarray}
\noindent Now our task is to find the range of $p$ for which $\widetilde{R}$ defines a positive map. 
 Based on the sign of $\lambda_{min}[R(\rho )]$, we consider the following two cases.\\
\textbf{Case-I:} When $\lambda_{min}[R(\rho )]\geq 0$, the RHS of the inequality in (\ref{lambmin}) is positive for every $0 \leq p\leq 1$ and hence $\widetilde{R}(\rho)$ represent a positive map for all $p \in [0,1]$.\\
\textbf{Case-II:} If $\lambda_{min}[R(\rho)]<0$, then (\ref{lambmin}) may be rewritten as 
\begin{eqnarray}
\lambda_{min}[\widetilde{R}(\rho)] &\geq& \frac{p}{d^2}+\frac{(1-p)}{Tr[R(\rho)]}\lambda_{min}^{lb}[R(\rho )]
\label{lambminneg}
\end{eqnarray}
where $\lambda_{min}^{lb}[R(\rho)]$ is given in (\ref{lb}) and may be re-expressed in terms of moments as
\begin{eqnarray}
\lambda_{min}^{lb}[R(\rho)]= \frac{m_1}{d^2} - \sqrt{(d^{2}-1)\left(\frac{m_2}{d^2}- \left(\frac{m_1}{d^2}\right)^2\right)}\nonumber\\ \label{lbr}
\end{eqnarray}
where $m_1 = Tr[R(\rho)]$ and $m_2 = Tr[(R(\rho))^2] $.\\
Taking $\lambda_{min}^{lb}[R(\rho)]=-k,~~k(>0)\in \mathbb{R}$, (\ref{lambminneg}) reduces to
\begin{eqnarray}
\lambda_{min}[\widetilde{R}(\rho)] &\geq&  \frac{p}{d^2}-k\frac{(1-p)}{Tr[R(\rho)]}
\label{lambminneg1}
\end{eqnarray}
Now, if we impose the condition on the parameter $p$ as $p\geq \frac{d^2k}{Tr[R(\rho)]+d^2k}=l$ then $\lambda_{min}[\widetilde{R}(\rho)]\geq 0$. 
Thus combining the above discussed two case, we can say that the approximated map $\widetilde{R}(\rho)$ represent a positive map when (\ref{thm1}) holds. Hence the theorem is proved.\\
\subsection{Completely positivity of structural physical approximation of realignment map}
In order to show that the approximated map $\widetilde{R}(\rho)$ defined in (\ref{sparealign}) may be realized in an experiment, it is not enough to show that $\widetilde{R}(\rho)$ is positive but also we need to show that it is completely positive.\\ 
When $l \leq p \leq 1$ there exist non-negative real numbers $\gamma_1$ and $\gamma_2$ such that the following conditions hold 
\begin{eqnarray}
\lambda_{min} [\widetilde{R}(\rho)] &\geq& \gamma_1 \lambda_{min}[\rho] \label{26}\\
\lambda_{max} [\widetilde{R}(\rho)] &\leq& \gamma_2 \lambda_{max}[\rho] \label{27}
\end{eqnarray}
Hence, using Result-3, $\widetilde{R}(\rho)$ is a completely positive operator for $p \in [l,1]$.
\section{Detection using the Experimental Implementable form of Realignment criteria}
\noindent In this section, we will derive a separability condition for the detection of NPTES and PPTES that may be implemented in the laboratory. The separability condition obtained depends on the structural physical approximation of Realignment criterion and thus the condition may be termed as SPA-R criterion. We will then further identify a class of states known as Schmidt-symmetric state for which the SPA-R criterion is equivalent to original form of realignment criterion \cite{rudolph2004,chen2} and weak form of realignment criterion \cite{hertz}. 
\subsection{SPA-R Criterion}
We are now in a position to derive the laboratory-friendly (for clarification, see Appendix-II) separability criterion that may detect the NPTES and PPTES. The proposed entanglement detection criterion is based on the structural physical approximation of Realignment criterion and it may be stated in the following  theorem.\\
\textbf{Theorem 2:} If any quantum system described by a density operator $\rho_{sep}$ in $d\otimes d$ system is separable then
\begin{eqnarray}
||\widetilde{R}(\rho_{sep})||_1 \leq \frac{p[Tr[R(\rho_{sep})]-1]+1}{Tr[R(\rho_{sep})]}=\widetilde{R}(\rho_{sep})_{UB} \label{thm2}
\end{eqnarray}
\textbf{Proof:} Let us consider a two-qudit bipartite separable state described by the density matrix $\rho_{sep}$, then the approximated realignment map (\ref{sparealign}) may be recalled as
\begin{eqnarray}
\widetilde{R}(\rho_{sep}) = \frac{p}{d^2}I_{d\otimes d}+\frac{1-p}{Tr[R(\rho_{sep})]}R(\rho_{sep})
\label{spa1}
\end{eqnarray}
Taking trace norm on both sides of (\ref{spa1}) and using triangular inequality on norm, it reduces to 
\begin{eqnarray}
||\widetilde{R}(\rho_{sep})||_1 &\leq& ||\frac{p}{d^2}I_{d\otimes d}||_1+||\frac{1-p}{Tr[R(\rho_{sep})]}R(\rho_{sep})||_1\nonumber\\
&=& p+\frac{1-p}{Tr[R(\rho_{sep})]}||R(\rho_{sep})||_1
\label{spa3}
\end{eqnarray}
Since $\rho_{sep}$ denote a separable state so using realignment criteria, we have $||R(\rho_{sep})||_1\leq 1$ \cite{chen,rudolph5}. Therefore, (\ref{spa3}) further reduces to
\begin{eqnarray}
||\widetilde{R}(\rho_{sep})||_1 &\leq& p+\frac{1-p}{Tr[R(\rho_{sep})]}\nonumber\\
&=& \frac{p[Tr[R(\rho_{sep})]-1]+1}{Tr[R(\rho_{sep})]}
\end{eqnarray}
Hence proved.\\
\textbf{Corollary-1:} If for any two-qudit bipartite state $\rho$, the inequality
\begin{eqnarray}
||\widetilde{R}(\rho)||_1 > \frac{p[Tr[R(\rho)]-1]+1}{Tr[R(\rho)]}=\widetilde{R}(\rho)_{UB} \label{cor1}
\end{eqnarray}   
holds then the state $\rho$ is an entangled state.\\
We should note an important fact that the $\widetilde{R}(\rho_{sep})_{UB}$ given in (\ref{thm2}) and (\ref{cor1}) depends on $Tr(R(\rho))$, which can be considered as the first moment of $R(\rho)$ and it may be measured in an experiment \cite{sougato} (See Appendix II).\\  
\textbf{Corollary-2:} If for any separable state $\rho_{sep}^{(1)}$, $Tr[R(\rho_{sep}^{(1)})]=1$ holds then (\ref{thm2}) reduces to 
\begin{eqnarray}
||\widetilde{R}(\rho_{sep}^{(1)})||_1 \leq 1
\label{cor2}
\end{eqnarray}
\subsection{Schmidt-symmetric states}
Let us consider a class of states known as Schmidt-symmetric states which may be defined as \cite{hertz} 
\begin{eqnarray}
\rho_{sc}=\sum_{i} \lambda_{i} A_{i}\otimes A_{i}^{*}
\label{sc}
\end{eqnarray}
where $A_{i}$ represent the orthonormal bases of the operator space and $\lambda_{i}$ denote non-negative real numbers known as Schmidt coefficients.\\ 
We are considering this particular class of states because we will show in this section that the separability criteria using SPA-R map becomes equivalent to the original form of realignment criteria for such class of state. Hertz et.al. studied the Schmidt-symmetric states and proved that a bipartite state $\rho_{sc}$ is Schmidt-symmetric if and only if 
\begin{eqnarray}
	||R(\rho_{sc})||_1 = Tr[R(\rho_{sc})] \label{sch}
\end{eqnarray}
For any Schmidt-symmetric state described by the density operator $\rho_{sc}$, the realignment matrix $R(\rho_{sc})$ defines a positive semi-definite matrix. Hence, using $Theorem-1$, $\widetilde{R}(\rho_{sc})$ is positive $ \forall \; p\in [0,1]$. Also, using (\ref{26}) and (\ref{27}), $\widetilde{R}(\rho_{sc})$ can be shown as a completely positive. To achieve the motivation of this section, let us start with the following lemma.\\
\textbf{Lemma 1:} For any Schmidt-symmetric state $\rho_{sc}$,
\begin{eqnarray}
||\widetilde{R}(\rho_{sc})||_1 = 1 
\label{lemma1}
\end{eqnarray}
\textbf{Proof:} Let us recall (\ref{sparealign}), which may provide the structural physical approximation of the realignment of Schmidt-symmetric state. The SPA-R of $\rho_{sc}$ is denoted by $\widetilde{R}(\rho_{sc})$ and it is given by
\begin{eqnarray}
	\widetilde{R}(\rho_{sc}) = \frac{p}{d^2}I_{d\otimes d}+\frac{(1-p)}{Tr[R(\rho_{sc})]}{R(\rho_{sc})}
\end{eqnarray}
Taking trace norm on both sides and using triangle inequality we have,
\begin{eqnarray}
||\widetilde{R}(\rho_{sc})||_1 \leq p + \frac{(1-p)}{Tr[R(\rho_{sc})]} ||R(\rho_{sc}) ||_1
\label{m1}
\end{eqnarray}
Using (\ref{sch}), the inequality (\ref{m1}) reduces to
\begin{eqnarray}
||\widetilde{R}(\rho_{sc})||_1 \leq 1
\label{r1}
\end{eqnarray}
Again, using (\ref{sparealign}), the trace of the approximated map $\widetilde{R}(\rho_{sc})$ is given by
\begin{eqnarray}
Tr[\widetilde{R}(\rho_{sc})] = Tr[\frac{p}{d^2}I_{d\otimes d}+\frac{(1-p)}{Tr[R(\rho_{sc})]}{R(\rho_{sc})}] = 1 
\label{trace}
\end{eqnarray}
Moreover, it is known that the trace norm of an operator is greater than or equal to its trace. Therefore, applying this result on $\widetilde{R}(\rho_{sc})$, we get
\begin{eqnarray}
Tr[\widetilde{R}(\rho_{sc})] \leq ||\widetilde{R}(\rho_{sc})||_1
\label{r11}
\end{eqnarray}
Using (\ref{trace}), the inequality (\ref{r11}) reduces to
\begin{eqnarray}
||\widetilde{R}(\rho_{sc})||_1 \geq 1
\label{ineq1}
\end{eqnarray}
Both (\ref{r1}) and (\ref{ineq1}) holds only when
\begin{eqnarray}
||\widetilde{R}(\rho_{sc})||_1 = 1
\label{r2}
\end{eqnarray}
holds. Thus proved.\\
We are now in a position to show that SPA-R criteria may reduce to original form of realignment criteria for Schmidt-symmetric states. It may be expressed in the following theorem.\\
\textbf{Theorem 3:} For Schmidt-symmetric state, SPA-R separability criterion reduces to the original form of realignment criterion.\\
\textbf{Proof:} Let $\rho_{sc}^{sep}$ be any separable Schmidt-symmetric state. The SPA-R separability criterion for $\rho_{sc}^{sep}$ is given by
\begin{eqnarray}
||\widetilde{R}(\rho_{sc}^{sep})||_1 \leq \widetilde{R}(\rho_{sc}^{sep})_{UB}
\label{f1} 
\end{eqnarray}
Using (\ref{lemma1}), the inequality (\ref{f1}) reduces to 
\begin{eqnarray}
&&\widetilde{R}(\rho_{sc}^{sep})_{UB} = \frac{p[Tr[R(\rho_{sc}^{sep})]-1]+1}{Tr[R(\rho_{sc}^{sep})]} \geq 1 \nonumber\\
&\implies& \;\; p[Tr[R(\rho_{sc}^{sep})]-1] + 1 \geq Tr[R(\rho_{sc}^{sep})]\nonumber\\
&\implies&\;\; Tr[R(\rho_{sc}^{sep})] (p-1) \geq (p-1)\nonumber\\
&\implies&\;\; Tr[R(\rho_{sc}^{sep})] \leq 1 \nonumber\\
&\implies&\;\; ||R(\rho_{sc}^{sep})||_1 \leq 1
\end{eqnarray}
The last step follows from (\ref{sch}). Hence the theorem.
\section{Error in the approximated map}
\noindent In this section, we have studied and analysed the error generated when $R(\rho)$ is approximated by its SPA. 
In the approximated map, we have added an appropriate proportion of maximally mixed state such that the approximated map has no negative eigenvalue. The error between the approximated map $\widetilde{R}(\rho)$ and the realignment map $R(\rho)$ may be calculated as:
\begin{eqnarray}
||\widetilde{R}(\rho)-R(\rho)||_1&=&||\frac{p}{d^2}I_{d\otimes d}+\frac{(1-p)}{Tr[R(\rho)]}R(\rho)-R(\rho)||_1\nonumber\\
&=&||\frac{p}{d^2}I_{d\otimes d}+[\frac{(1-p)}{Tr[R(\rho)]}-1]R(\rho)||_1\nonumber\\
\label{rrho1}
\end{eqnarray}
Using triangular inequality for trace norm, (\ref{rrho1}) reduces to
\begin{eqnarray}
||\widetilde{R}(\rho)-R(\rho)||_1\leq p+{\frac{1-p-Tr[R(\rho)]}{Tr[R(\rho)]}}||R(\rho)||_1
\label{error}
\end{eqnarray}
The inequality (\ref{error}) may be termed as error inequality. The error inequality holds for any two-qudit bipartite state.\\
\textbf{Proposition 1:} The equality relation   
\begin{eqnarray}
||\widetilde{R}(\rho_{sep})-R(\rho_{sep})||_1= \frac{(1-p)(1-Tr[R(\rho_{sep})])}{Tr[R(\rho_{sep})]}
	\label{errorsep1}
\end{eqnarray}
holds for separable state described by the density operator $\rho_{sep}$ such that $||R(\rho_{sep})||_1=1$.\\
\textbf{Proof:} Equality in (\ref{error}) holds if and only if 
\begin{eqnarray}
	\frac{p}{d^2}I_{d\otimes d}=[\frac{(1-p)}{Tr[R(\rho)]}-1]R(\rho)
\end{eqnarray}
i.e. equality in (\ref{error}) holds when the realigned matrix takes the form
\begin{eqnarray}
	R(\rho)=\frac{pTr[R(\rho)]}{1-p-Tr[R(\rho)]}\frac{I}{d^2},~~0\leq p\leq 1
	\label{eq22}
\end{eqnarray} 
Taking trace norm, (\ref{eq22}) reduces to
\begin{eqnarray}
||R(\rho)||_1=\frac{pTr[R(\rho)]}{1-p-Tr[R(\rho)]},~~0\leq p\leq 1
\label{eq23}
\end{eqnarray}
For separable state $\rho_{sep}$, (\ref{eq23}) reduces to
\begin{eqnarray}
1=\frac{pTr[R(\rho)]}{1-p-Tr[R(\rho)]},~~0\leq p\leq 1
\label{eq24}
\end{eqnarray}
Simplifying (\ref{eq24}), the value of $p$ and $1-p$ may be expressed as
\begin{eqnarray}
	p=\frac{1-Tr[R(\rho_{sep})]}{1+Tr[R(\rho_{sep})]},~~1-p=\frac{2Tr[R(\rho_{sep})]}{1+Tr[R(\rho_{sep})]}
\end{eqnarray}
Substituting values of $p$ and $1-p$ in (\ref{eq22}), the realigned matrix for separable state, $R(\rho_{sep})$ takes the form
\begin{eqnarray}
	R(\rho_{sep})=\frac{1}{d^2}I
	\label{mes}
\end{eqnarray}
Therefore, (\ref{mes}) holds only for separable states. This means that there exist a separable state $\rho_{sep}$ such that $||\rho_{sep}||_{1}=1$ for which the equality condition in the error inequality (\ref{error}) holds.\\

\textbf{Result-4} If any quantum system described by a density operator $\rho$ in $d\otimes d$ system is separable then the error inequality is given by
\begin{eqnarray}
	||\widetilde{R}(\rho)-R(\rho)||_1&\leq&  \frac{(1-p)[1-Tr[R(\rho)]]}{Tr[R(\rho)]}
	\label{errorsep1}
\end{eqnarray}
\textbf{Proof:} Let us consider a separable state $\rho_{sep}$. Thus, we have $||R(\rho_{sep})||_{1}\leq 1$. Therefore, error inequality (\ref{error}) reduces to
\begin{eqnarray}
||\widetilde{R}(\rho_{sep})-R(\rho_{sep})||_1&\leq& p+{\frac{1-p-Tr[R(\rho_{sep})]}{Tr[R(\rho_{sep})]}}\nonumber\\
&=& \frac{(1-p)[1-Tr[R(\rho_{sep})]]}{Tr[R(\rho_{sep})]}
\label{errorsep}
\end{eqnarray}
Hence proved.\\
\textbf{Corollary 3:} If inequality (\ref{errorsep1}) is violated by any bipartite $d \otimes d$ dimensional quantum state, then the state under investigation is entangled.\\
\section{Illustrations}
	\textbf{Example 1:} Consider the family of two-qubit states $\rho(r,s,t)$ discussed in \cite{rudolph3}. For $r=\frac{1}{4}$ and $s=\frac{1}{2}$, the family is represented by
	\begin{eqnarray}
	\rho_t=\frac{1}{2}
	\begin{pmatrix}
	\frac{5}{4} & 0 & 0 & t\\
	0 & 0 & 0 & 0\\
	0 & 0 & \frac{1}{4} & 0\\
	t & 0 & 0 & \frac{1}{2}
	\end{pmatrix} \label{rhot} 
	\end{eqnarray}
$\rho_{t}$ may be defined as a valid quantum state when $|t| \leq \frac{\sqrt{\frac{5}{2}}}{2} \approx 0.790569$. By PPT criterion, $\rho_t$ is entangled when $t \neq 0$. Realignment criteria detect the entangled states for $|t| > 0.116117$.\\
Using the prescription given in (\ref{sparealign}), we construct the SPA-R map $\widetilde{R}: M_4 (\mathbb{C}) \longrightarrow M_4 (\mathbb{C})$ as
\begin{eqnarray}
	\widetilde{R}(\rho_t) = \frac{p}{4} I_4 + \frac{(1-p)}{Tr[R(\rho_t)]} R(\rho_t)
\end{eqnarray}
where $0 \leq p \leq 1$.\\
Using Descarte's rule of sign, we find that $R(\rho_t)$ is positive semi-definite for $t \geq 0$ (Detailed calculation given in Appendix).\\
Applying $Theorem-1$, it can be shown that the approximated map $\widetilde{R}(\rho_t)$ is positive as well as completely positive for $l \leq p \leq 1$ where
	\begin{eqnarray}
		l = 
		\left\{
		\begin{array}{lrr}
		p_1(t) &\text{if }& -0.790569 \leq t < 0\\
			0 & \text{if }& 0 \leq t \leq 0.790569
		\end{array}
	\right\}
	\end{eqnarray}
	where 
\begin{eqnarray}
	p_1(t) = 	\frac{2(13-24t+8t^2)-\sqrt{3(67-112t+64t^2)}}{(-5+4t)^2} \label{p1}
\end{eqnarray}
Thus, the SPA-R map $\widetilde{R}(\rho_t)$, which is a completely positive map may be suitable for detecting the entanglement in the family of states described by the density operator $\rho_{t}$.
Now we apply our separability criterion discussed in $Theorem-2$ which involves the comparison of  $||\widetilde{R}(\rho_t)||_1$ and the upper bound $\widetilde{R}(\rho_t)_{UB}$ defined in (\ref{thm2}).
After few steps of the simple calculation, we obtain
	\begin{eqnarray}
||\widetilde{R}(\rho_t)||_1 > \widetilde{R}(\rho_t)_{UB} \label{rhoteq} 
\end{eqnarray}
for\\
$\begin{array}{lrl}
		 t \in (-0.790569, -0.665506] &\text{when}& p_1(t) \leq p < p_2(t) \\
		 t \in (0.116117, 0.125] &\text{when}& 0 \leq p < p_3(t) \\
		 t \in (0.125, 0.790569] &\text{when}& 0 \leq p\leq 1
\end{array}$

where 
\begin{eqnarray}
	p_2(t) &=& \frac{(-91 - 48 t - 64 t^2) -
		\sqrt{u(t)}}{2
		(-7 + 48 t)^2}\\
	p_3(t) &=&  \frac{(14 - 128 t + 64 t^2)}{(7 - 80 t + 128 t^2)}
\end{eqnarray}
The function $u(t)$ is given by $u= 8673 + 9632 t - 8832 t^2 - 6144 t^3 + 4096 t^4$.\\
Thus, the inequality (\ref{thm2}) is violated when $t > 0.116117$ and $t<-0.665506$ which implies that the state $\rho_{t}$ is entangled for $t \in [-0.790569,-0.665506) \cup (0.116117, 0.790569] $.\\
The comparison of $||\widetilde{R}(\rho_t)||_1$ and $\widetilde{R}(\rho_t)_{UB}$ for the two-qubit state $\rho_t$ has been studied in Fig-1 for different range of $t$ given in (\ref{rhoteq}). From Fig-1 it can be observed that the inequality (\ref{rhoteq}) holds for $t > 0.116117$ which implies that the entanglement of $\rho_t$ is detected in this region.
\begin{figure}[h]
	\centering
	\includegraphics[scale=.4]{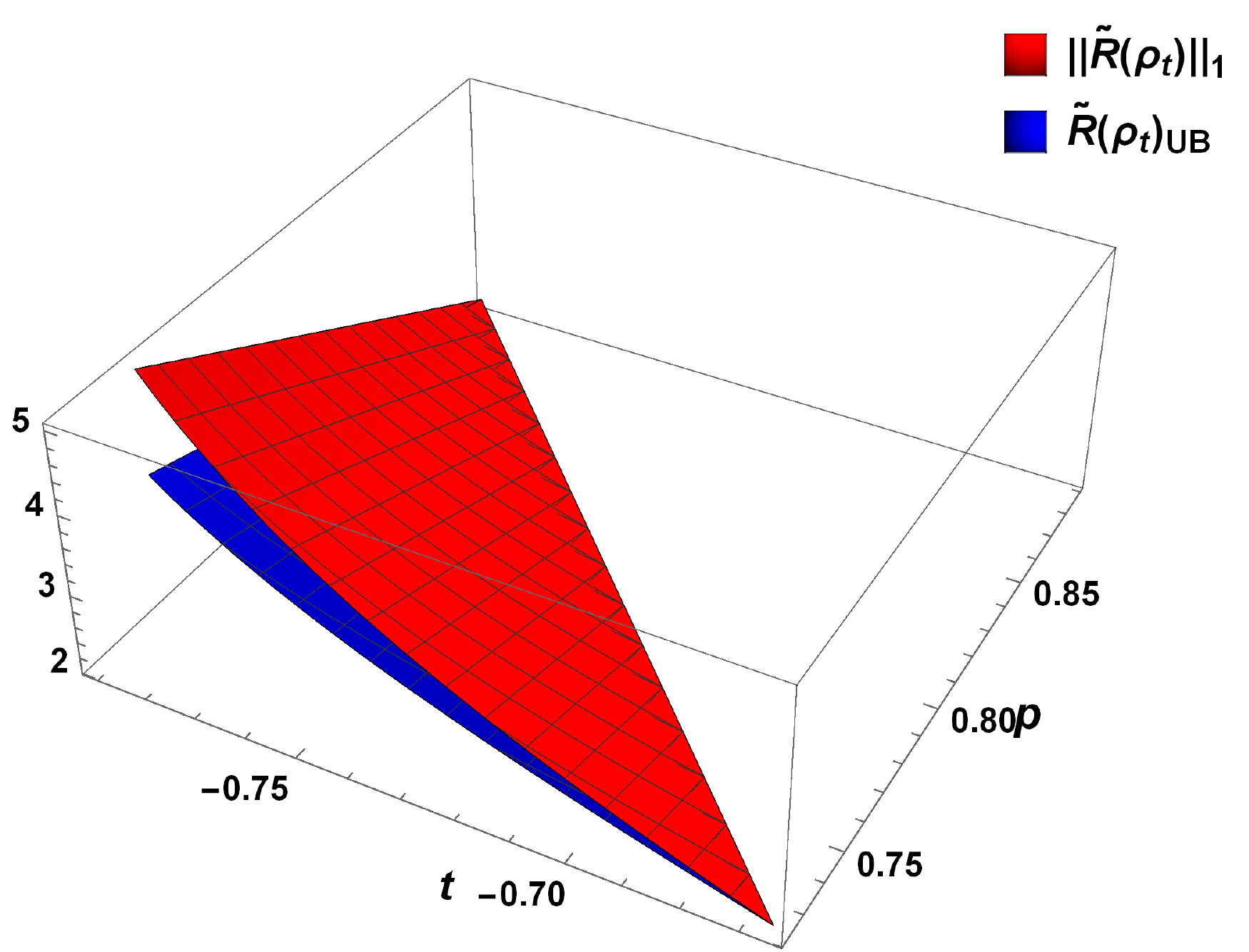}
	1a.
	\includegraphics[scale=.55]{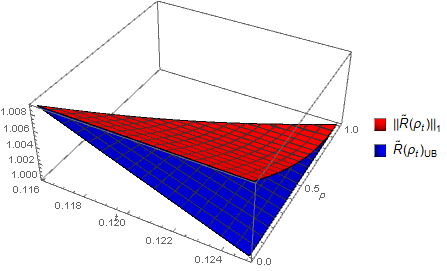}
	1b.
	\includegraphics[scale=.55]{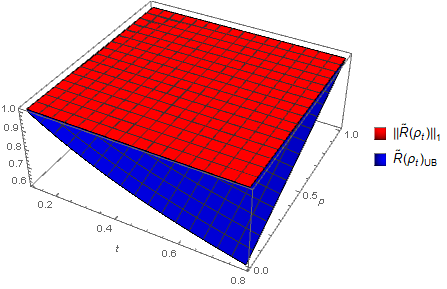}
	1c.
	\caption{The comparison between the $||\widetilde{R}(\rho_t)||_1$ and $\widetilde{R}(\rho_t)_{UB}$ for the two-qutrit state $\rho_{t>0}$ has been displayed. In Fig 1a., one can observe that the inequality (\ref{thm2}) obtained in Theorem-2 is violated when $-0.790569 \leq t < -0.665506$ for $p \in [p_1(t),p_2(t)]$ whereas in Fig 1b. the inequality is violated when $0.116117< t \leq 0.125$ and $p$ lies in the interval $[0,p_3(t))$. Fig. 1c. shows the violation of inequality (\ref{thm2}) when $t > 0.125$ and $0 \leq p \leq 1$.}
\end{figure}\\

\textbf{Example 2:} Consider a two-qutrit state defined in \cite{swapan}, which is described by the density operator
\begin{eqnarray}
	\rho_a = \frac{1}{5+2a^2}\sum_{i=1}^{3}{|\psi_i\rangle \langle \psi_i|},~~\frac{1}{\sqrt{2}}\leq a \leq 1
	\label{eg2}
\end{eqnarray}
where, $|\psi_i\rangle=|0i\rangle-a|i0\rangle$, for $i=\{1,2\}$ and\\ $|\psi_3\rangle=\sum_{i=0}^{2}{|ii\rangle}$.\\
The state described by the density operator $\rho_a$ is NPTES \cite{swapan}.
Using the prescription given in (\ref{sparealign}), we construct the SPA-R map $\widetilde{R}: M_9 (\mathbb{C}) \longrightarrow M_9 (\mathbb{C})$ as
\begin{eqnarray}
	\widetilde{R}(\rho_a) = \frac{p}{9} I_9 + \frac{(1-p)}{Tr[R(\rho_a)]}{R(\rho_a)},~ 0 \leq p \leq 1
\end{eqnarray}
Using Descarte's rule of sign, we find that $R(\rho_a)$ is not a positive semi-definite operator. (Detailed calculation given in Appendix).
Using $Theorem-1$, the approximated map $\widetilde{R}(\rho_a)$ is positive as well as completely positive for $l_{1} \leq p \leq 1$ where
\begin{eqnarray}
	l_{1} = \frac{-1+15\sqrt{2}w+6\sqrt{2}a^2w}{3\sqrt{2}(5+2a^2)w},~w=\sqrt{\frac{1}{56+9a^2(5+a^2)}}
\end{eqnarray}
Thus, the SPA-R map $\widetilde{R}(\rho_a)$ is suitable for detecting the entanglement in the state $\rho_a$ experimentally.
Now we apply our separability criterion discussed in $Theorem-2$ which involves the comparison of  $||\widetilde{R}(\rho_a)||_1$ and the upper bound $\widetilde{R}(\rho_a)_{UB}$ defined in (\ref{thm2}).
For $\frac{1}{\sqrt{2}}\leq a \leq 1$, we find that
\begin{eqnarray}
	||\widetilde{R}(\rho_a)||_1 > \widetilde{R}(\rho_a)_{UB}
\end{eqnarray}
The comparison of $||\widetilde{R}(\rho_a)||_1$ and $\widetilde{R}(\rho_a)_{UB}$ for the two-qutrit state $\rho_a$ has been studied in Fig-2.
From Fig-2, it is evident that the inequality (\ref{thm2}) obtained in $Theorem-2$ is violated. Thus, the state described by the density operator $\rho_a$ is an entangled state.\\
\begin{figure}[h]
	\centering
	\includegraphics[scale=.38]{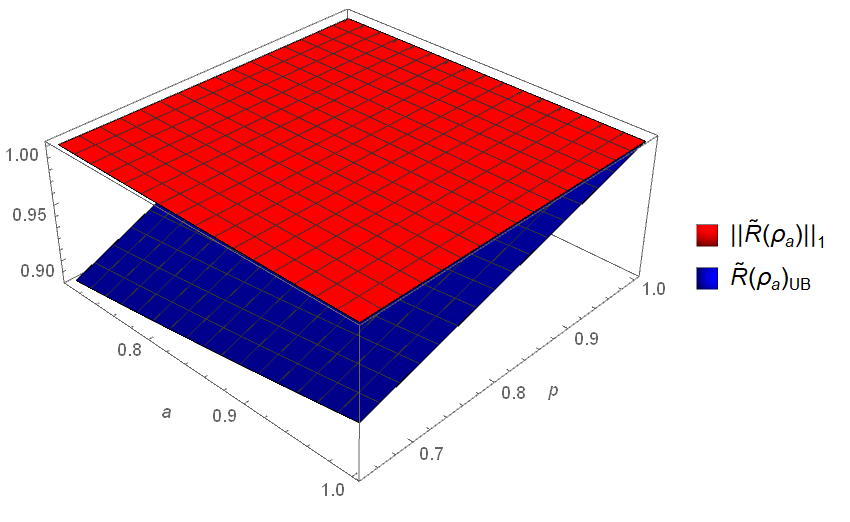}
	\caption{The comparison between the $||\widetilde{R}(\rho_a)||_1$ and $\widetilde{R}(\rho_a)_{UB}$ for the two-qutrit state $\rho_a$ has been displayed. It has been observed that the inequality (\ref{thm2}) is violated for $\rho_a$ in the whole range of $a$ and for $p \in [l_{1}, 1]$}
\end{figure}

\textbf{Example 3:} Let us consider a two-qutrit isotropic state described by the density operator $\rho_{\beta}$ \cite{iso}
\begin{eqnarray}
	\rho_{\beta}=\beta|\phi_{+}\rangle \langle \phi_{+}|+\frac{1-\beta}{9}I_9, ~~-\frac{1}{8}\leq \beta \leq 1
	\label{betastate}
\end{eqnarray}
where $I_9$ denotes the identity matrix of order 9 and the state $|\phi_{+}\rangle$ represents a Bell state in a two-qutrit system and may be expressed as
\begin{eqnarray}
	|\phi_{+}\rangle=\frac{1}{\sqrt{3}}(|11\rangle +|22\rangle +|33\rangle)
	\label{bellstate}
\end{eqnarray}
Using realignment criteria, the state $\rho_{\beta}$ is an entangled state for $\frac{1}{3}< \beta \leq 1$. Using Descarte's rule of sign, we find that the realignment matrix $R(\rho_{\beta})$ is positive semi-definite. The comparison between $||\widetilde{R}(\rho_{\beta})||_1$ and $\widetilde{R}(\rho_{\beta})_{UB}$ has been studied in Fig-3. 
\begin{figure}[h]
	\centering
	\includegraphics[scale=.33]{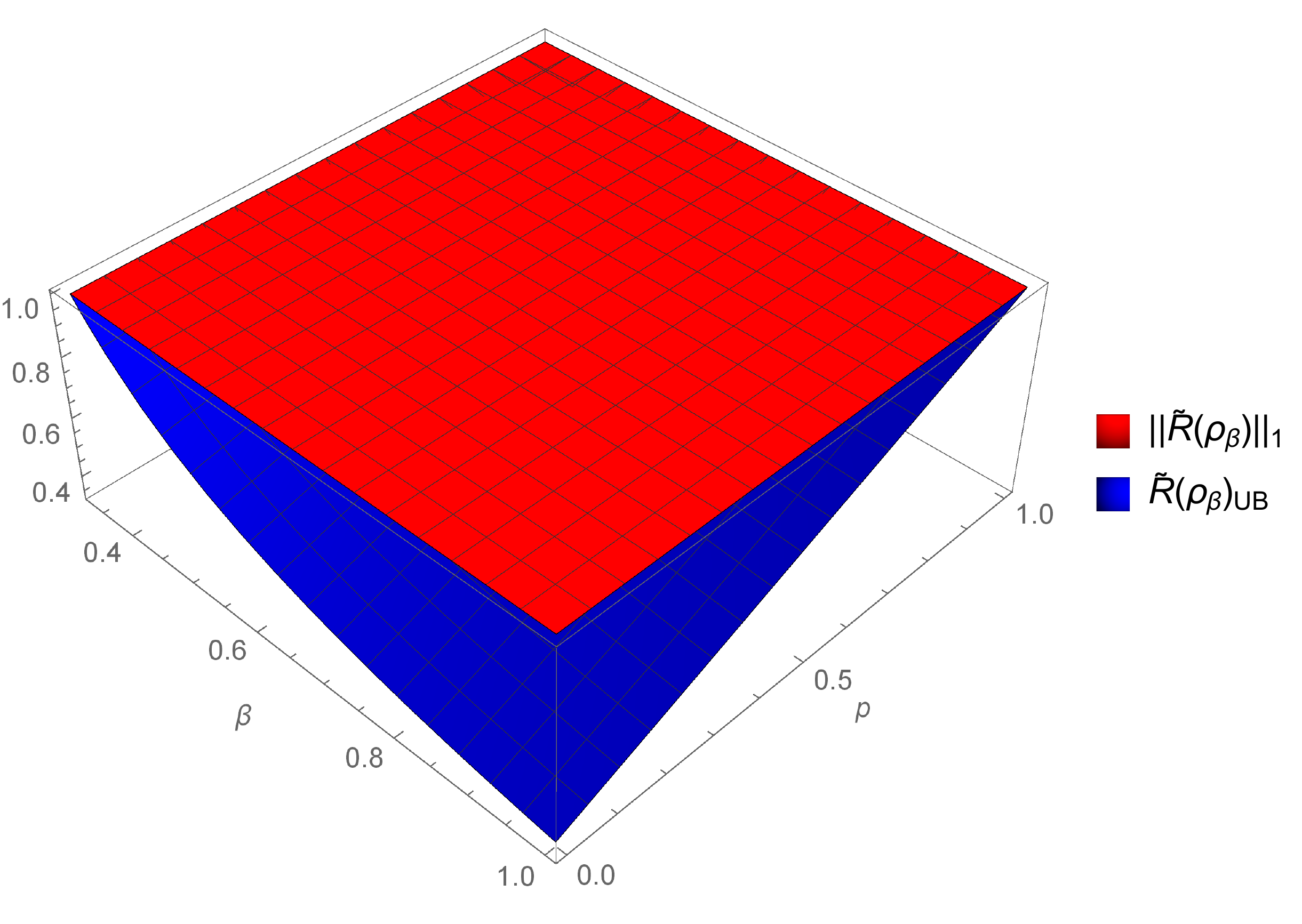}
	\caption{The comparison between the $||\widetilde{R}(\rho_{\beta})||_1$ and $\widetilde{R}(\rho_{\beta})_{UB}$ for the two-qutrit state $\rho_{\beta}$ has been displayed. It has been observed that the inequality (\ref{thm2}) is violated for all values of $\beta \in (1/3, 1]$ and for any $p \in [0,1]$.}
\end{figure}
From Fig-3, it can be observed that the inequality (\ref{thm2}) is violated for $\frac{1}{3}< \beta \leq 1$ and $p\in [0,1]$. Thus, using $Theorem-2$, the state described by the density operator $\rho_{\beta}$ is an entangled state.\\
	
	\textbf{Example 4:} Consider $\alpha$-state for $0\leq \alpha \leq 1$ described by the density operator $\rho_{\alpha}$ may be defined as 
	\begin{eqnarray}
	\rho_{\alpha}=\frac{1}{8\alpha+1}
	\begin{pmatrix}
	\alpha & 0 & 0 & 0 & \alpha & 0 & 0 & 0 & \alpha \\
	0 & \alpha & 0 & 0 & 0 & 0 & 0 & 0 & 0 \\
	0 & 0 & \alpha & 0 & 0 & 0 & 0 & 0 & 0 \\
	0 & 0 & 0 & \alpha & 0 & 0 & 0 & 0 & 0 \\
	\alpha & 0 & 0 & 0 & \alpha & 0 & 0 & 0 & \alpha \\
	0 & 0 & 0 & 0 & 0 & \alpha & 0 & 0 & 0 \\
	0 & 0 & 0 & 0 & 0 & 0 & \frac{1+\alpha}{2} & 0 & \frac{\sqrt{1-\alpha^2}}{2} \\
	0 & 0 & 0 & 0 & 0 & 0 & 0 & \alpha & 0 \\
	\alpha & 0 & 0 & 0 & \alpha & 0 & \frac{\sqrt{1-\alpha^2}}{2} & 0 & \frac{1+\alpha}{2} \\
	\end{pmatrix}
	\label{eg4}
	\end{eqnarray}\\
	It has been shown that this state is PPTES for $0<\alpha<1$ \cite{horodecki7}. Using Descarte's rule of sign, we find that the realignment matrix $R(\rho_{\alpha})$ is positive semi- definite (See appendix for detailed calculation). Further, using $Result-2$, it can be easily shown that the SPA-R map $\widetilde{R}(\rho_{\alpha})$ is completely positive for any $p \in[0,1]$. It has been observed that the inequality (\ref{thm2}) is violated for different range of $p$ for some values of $\alpha$, which is shown in the table given below.
	
	\begin{table}[h!]
		\begin{center}
			\begin{tabular}{ p{2.0cm} p{2.8cm}  p{1.8cm} }
				\hline
				$\alpha$   & $Range~of~p$   & $Theorem-2$  \\ 
				\hline
				$0.1$  &$0\leq p \leq 0.019383$  & $Violated$ \\
				$0.2$  &$0\leq p \leq 0.022143$  & $Violated$\\
				$0.3$  &$0\leq p \leq 0.021903$ & $Violated$\\
				$0.4$  &$0\leq p \leq 0.020444$ & $Violated$\\
				$0.5$  &$0\leq p \leq 0.018284$ & $Violated$\\
				$0.6$  &$0\leq p \leq 0.015611$ & $Violated$\\
				$0.7$  &$0\leq p \leq  0.012488$  & $Violated$\\
				$0.8$  &$0\leq p \leq 0.008904$  & $Violated$\\
				$0.9$  &$0\leq p \leq  0.004791$  & $Violated$\\
				\hline   
			\end{tabular}
		\end{center}
		\caption{The table shows the range of the probability $p$ for which the inequality (\ref{thm2}) is violated for different values of the state parameter $\alpha$}.
		\label{table1}
	\end{table}
Thus, we have shown that the criterion given by $Theorem-2$ is violated by $\rho_{\alpha}$ and thus our criterion detect the bound entangled state given by (\ref{eg4}).
\section{Efficiency of SPA-R criterion}
In this section, we show how SPA-R criterion is efficient in comparison to other entanglement detection criteria. In particular, we are considering three entaglement detection criterion such as (a) separability criterion based on realigned moment \cite{tzhang} and (b) partial realigned moment criterion \cite{shruti4} for comparing the efficiency of SPA-R criterion.
\subsection{Comparing SPA-R and moment based criterion (a)}
\begin{figure}[h!]
	\centering
	\includegraphics[scale=.48]{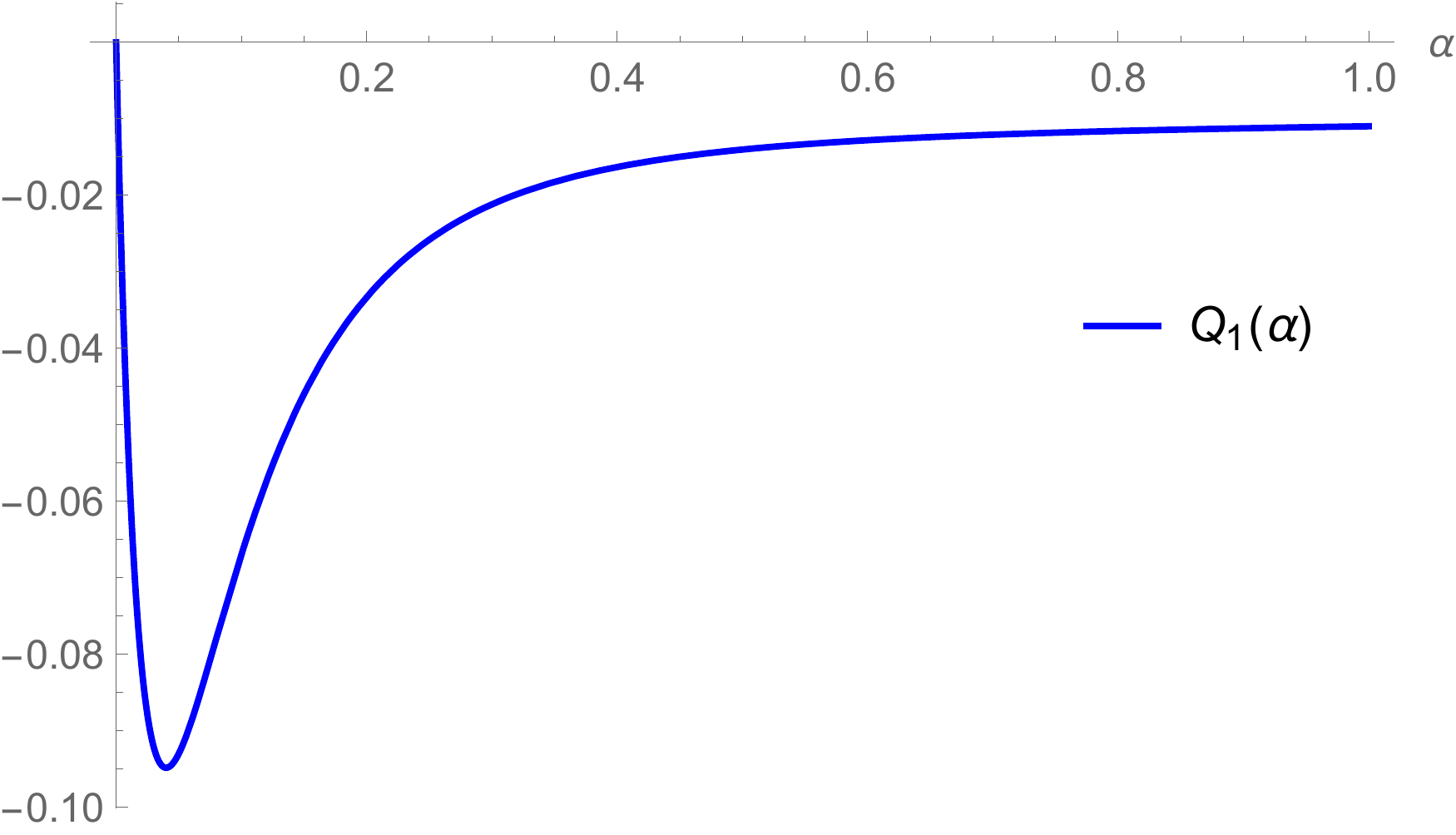}
	\caption{The red curve represents $Q_1$ for the state $\rho_{\alpha}$ and \textit{x}-axis depicts the state parameter $\alpha$.}.
	\label{q2al}
\end{figure}

To compare the SPA-R criterion with the moment based criterion (a), we will use example-1 and example-4.\\
(i) Let us recall example-1, where the family of states is described by the density operator $\rho_t$. Interestingly, for this family of states when $t>0$, our SPA-R criteria detects entanglement in the region $t \in (0.116117, 0.790569]$. But the realignment moment based criteria given in \cite{tzhang} detect the entangled state in the range $t\in (0.370992,  0.790569]$. Clearly, SPA-R criteria detects the NPTES $\rho_t$ for $t>0$ in a better range than the moment based criteria (a).\\
(ii) Let us consider the BES studied in example-4, which is described by the density operator $\rho_{\alpha}$, $0< \alpha <1$. The realignment moment for a bipartite state $\rho_{\alpha}$ may be defined as  \cite{tzhang}
\begin{eqnarray}
	r_k (R(\rho_{\alpha})) = Tr[R(\rho_{\alpha}) (R(\rho_{\alpha}))^{\dagger}]^{k/2},\; k = 1, 2, 3, . . ., n \label{zhangdef}
\end{eqnarray}
where $n$ denote the order of the matrix $R(\rho_{\alpha})$.\\
The separability criterion based on realignment moments $r_2$ and $r_3$ may be stated as \cite{tzhang}: If a quantum state $\rho_{\alpha}$ is separable, then
\begin{eqnarray}
	Q_1 = (r_2(R(\rho_{\alpha}))^2 - r_3(R(\rho_{\alpha}) \leq 0	\label{rzhang}
\end{eqnarray}
$Q_1 > 0$ certifies that the given state is entangled.\\
Fig-\ref{q2al} shows that the inequality (\ref{rzhang}) is not violated for the BES $\rho_{\alpha}$ in the whole range $0 < \alpha < 1$. Hence the BES $\rho_{\alpha}$ is undetected by this realignment moment based criteria.
\subsection{Comparing SPA-R and moment based criterion (b)}
\begin{figure}[h!]
	\centering
	\includegraphics[scale=.48]{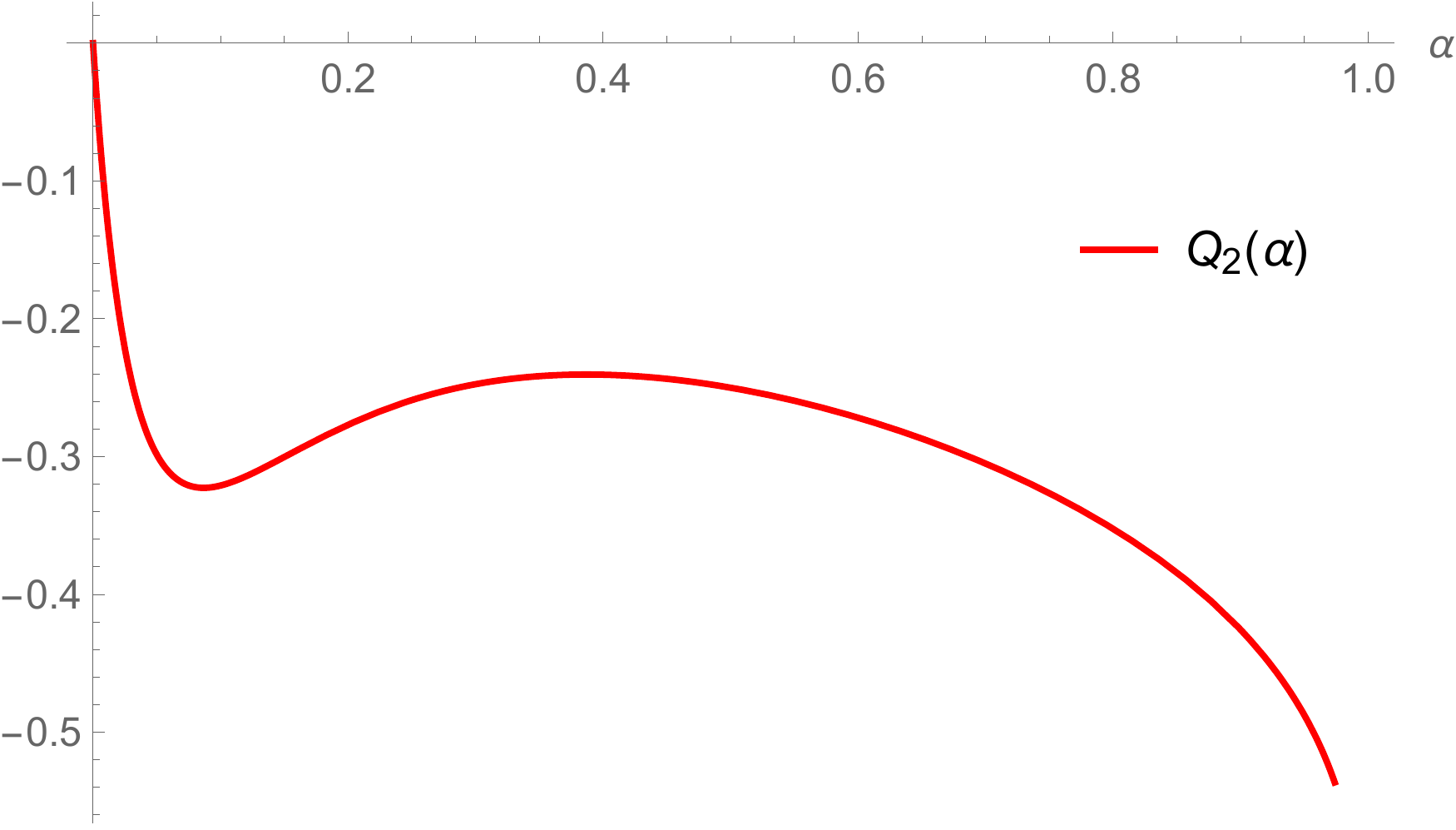}
	\caption{The red curve represents $Q_2$ for the state $\rho_{\alpha}$ and x-axis depicts the state parameter $\alpha$.}.
	\label{q1al}
\end{figure}

Let us again recall example-1 and example-4 to compare the SPA-R criterion with the moment based criterion (b).\\ 
(i) In the example-1, the family of states described by the density operator $\rho_t$. By $R$-moment criterion given in \cite{shruti4}, $\rho_t$ is detected when $t \in (0.214312, 0.790569] \subset (0.116117, 0.790569] $. Therefore, SPA-R criteria detects' more entangled states than the $R$-moment criterion.\\
(ii) Let us now consider the BES studied in example-4. Applying $R$-moment criterion \cite{shruti4} on the BES described by the density operator $\rho_{\alpha}$, $0<\alpha<1$, we get
 \begin{eqnarray}
Q_2 \equiv 	56 D_8^{1/8} + T_1 - 1 \leq 0 \;\; \forall \;  \alpha \in (0,1) \label{ineqal}
\end{eqnarray}
where $D_8 = \prod_{i=1}^8 \sigma_i^2(\rho_{\alpha})$ and $T_1 = Tr[R(\rho_{\alpha})]$. Here $\sigma_i(\rho_{\alpha})$ represents the $ith$ singular value of $\rho_{\alpha}$.
 Since the above inequality is not violated for any $\alpha$, the BES $\rho_{\alpha}$ is undetected by  $R$-moment based criteria. This is shown in Fig-\ref{q1al}.
\section{Conclusion}
\noindent To summarize, we have developed a separability criterion by approximating realignment operation via structural physical approximation (SPA). Since the partial transposition (PT) operation is limited to detect only NPTES, so we have studied here the realignment operation, which may detect both NPTES and PPTES. But since realignment map is not a positive map and thus it does not represent a completely positive map so it is difficult to implement it in a laboratory. Therefore, in order to make realignment map completely positive, firstly, we have approximated it to a positive map using the method of SPA and then we have shown that this approximated map is also completely positive. We have shown that the positivity of the SPA-R map can be verified in an experiment because the lower bound of the fraction $p$ can be expressed in terms of the first and second moments of the realignment matrix. Interestingly, we have shown that the separability criterion derived in this work using the approximated (SPA-R) map detect NPT and PPT bipartite entangled state.  Some examples are cited to support the result obtained in this work. Although there are other PPT criterion that may detect NPTES and PPTES but our result is interesting in the sense that it may be realized in an experiment. Our obtained results may be realized in an experiment but to achieve this aim, we pay a price in terms of the short range detection. This fact can be observed in Example 1 where the range of the state parameter for the detection of entangled state is smaller than the range obtained by realignment operation (without approximation). We also have analyzed the error occured during the structural physical approximation of the realignment map and it is described by an inequality known as error inequality. Lastly, we have obtained an inequality which is satisfied by all bipartite $d\otimes d$ dimensional separable state and the violation of this inequality guarantees the fact that the state under probe is entangled. Interestingly, the SPA-R criteria conincides with the original realignment criteria for Schmidt-symmetric states.\\
	
\section{Acknowledgement}
S. Aggarwal and A. Kumari would like to acknowledge the financial support from CSIR. This work is supported by CSIR File No. 08/133(0043)/2019-EMR-1 and 08/133(0027)/2018-EMR-1 respectively.\\

\section*{Appendix I}
\subsection{Example 1}
Consider the $2$-qubit state $\rho_t$ defined in (\ref{rhot}). The characteristic polynomial of the matrix $R(\rho_t)$ can be expressed as
\begin{eqnarray}
	f_1(x) = x^4 - a_1(t) x^3 + a_2(t) x^2 - a_3(t) x + a_4(t)
\end{eqnarray}
Using (\ref{rec}), we get
\begin{eqnarray}
	&&a_1(t) = m_1 = t + \frac{7}{8}
	\label{a1t}\\
	&&a_2(t) =\frac{1}{2} (m_1^2 - m_2) = \frac{1}{32}(8t^2 +28 t + 5)\\
	&&a_3(t) = \frac{1}{6} ( m_1^3 - 3 m_1 m_2 + 2m_3)= \frac{1}{32}(7t^2 + 5t)\\
	&&a_4(t) = \frac{1}{24} (m_1^4 - 6 m_1^2 m_2 + 8 m_1 m_3 + 3 m_2^2 - 6m_4) = \frac{5}{128}t^2 \nonumber\\
\end{eqnarray}
where $m_k=Tr[(R(\rho_t))^k]$

$R(\rho_t)$ is positive semi-definite iff $a_i(t) \geq 0$ for all $i = 1$ to $4$.
After simple calculation we get, 
\begin{eqnarray*}
	&&a_1(t) > 0 \;\; \text{for} \;\; t \in [-0.790569, 0.790569] \\
	&&a_2(t) \geq 0 \;\; \text{for} \;\; t \in [-0.188751, 0.790569]\\
	&&a_3(t) \geq 0 \;\; \text{for} \;\; t \in [-0.790569, -0.714286] \cup [0, 0.790569] \nonumber\\
	&&a_4(t) \geq 0 \;\; \text{for} \;\; t \in [-0.790569, 0.790569]
\end{eqnarray*}
From above calculations, we observe the following:\\
\textbf{Case 1:} If $t\geq 0$ then all the coefficients of the characteristic polynomial $f_1(-x)$ are positive, i.e., there is no sign change in the ordered list of coefficients of $f_1(-x)$. Thus $R(\rho_t)$ has no negative eigenvalue for $t\geq 0$.\\
\textbf{Case 2:} If $t < 0$ then (i) $a_2(t) < 0$ for $t \in [-0.790569, -0.188751]$ and (ii) $a_3(t) < 0$ for $t \in [-0.714286, 0] $, Hence, for every $t$, atleast one coefficient of $f_1(x)$ is negative. Hence $R(\rho_t)$ has atleast one negative eigenvalue, i.e.,  $R(\rho_t)$ is not positive semi-definite (PSD) for $t<0$.\\
From above analysis, it is trivial that $\widetilde{R}(\rho_{t})$ defines a positive map when $t \geq 0$. Now by $Theorem-1$, for $t < 0$, $\widetilde{R}(\rho_{t}) > 0$ when the lower bound $l$ of the proportion $p$ is given as
\begin{eqnarray}
	l = \frac{4k}{Tr[R(\rho_t)] + 4k} = p_1(t)
\end{eqnarray}
where $Tr[R(\rho_t)]=a_{1}(t)$ is given in (\ref{a1t}), $p_1 (t)$ is defined in (\ref{p1}) and $k$ is given by
\begin{eqnarray}
	k &=& -\lambda_{min}^{lb}[\rho_t] \nonumber\\&=& \frac{-1}{32} (7+ 8t - \sqrt{3(67 - 112 t + 64 t^2)})
\end{eqnarray}
\subsection{Example 2}
Consider the two-qutrit state defined in (\ref{eg2}).

Let $f_2(x)$ be chracteristic polynomial of $\rho_{a}$ given as
\begin{eqnarray}
	f_2(x) &=& x^9 - a_1(a) x^8 +a_2(a) x^7 - a_3(a) x^6 + a_4(a) x^5 \nonumber \\&& - a_5(a) x^4  + a_6(a) x^3 - a_7(a) x^2 + a_8(a) x - a_{9}(a)\nonumber
\end{eqnarray}
where the coefficients $a_i(a)$ calculated in terms of moments using (\ref{rec}) are given as
\begin{eqnarray}
	a_1(a)&=&\frac{9}{5+2a^2},\nonumber\\a_2(a) &=&- \frac{4(-9+a^2)}{(5+2a^2)^2}, \nonumber\\ a_3(a) &=& -\frac{28(-3+a^2)}{(5+2a^2)^3},\nonumber\\a_4(a) &=& \frac{126-84a^2+5a^4}{(5+2a^2)^4}, \nonumber\\ a_5(a) &=& \frac{126-140a^2+25a^4}{(5+2a^2)^5},  \; \;\nonumber\\ a_6(a) &=& -\frac{2(-42+70a^2-25a^4+a^6)}{(5+2a^2)^6}, \nonumber\\
	a_7(a) &=&-\frac{2(-18+42a^2-25a^4+3a^6)}{(5+2a^2)^7},\nonumber\\ a_8(a) &=&- \frac{-9+28a^2-25a^4+6a^6}{(5+2a^2)^8}
\end{eqnarray}
and 
\begin{eqnarray}
	a_9(a) =-\frac{(-1+a^2)^2(-1+2a^2)}{(5+2a^2)^9}
\end{eqnarray}.
From the coefficients of $f_2(x)$, it can be observed that atleast one coefficient of $f_2(x)$ is negative. This means $R(\rho_a)$ has atleast one negative eigenvalue, i.e.,  $R(\rho_a)$ is not PSD.

Using $Theorem-1$, the approximated map $\widetilde{R}(\rho_a)$ is positive as well as completely positive when the lower bound $l$ of the proportion $p$ is given as
\begin{eqnarray}
	l = \frac{9k}{Tr[R(\rho_a)] + 9k}
\end{eqnarray}
where $Tr[R(\rho_a)]$ is the trace of $R(\rho_a)$ and
\begin{eqnarray}
	k &=& -\lambda_{min}^{lb}[\rho_a] \nonumber\\&=& -\frac{1}{5+2a^2}+3\sqrt{2}\sqrt{\frac{1}{56+45a^2+9a^4}}
\end{eqnarray}
Substituting value of $k$ and $Tr[R(\rho_a)]$, the lower bound $l_1$ may be expressed as
\begin{eqnarray}
	l_1 = \frac{-1+15\sqrt{2}w+6\sqrt{2}a^2w}{3\sqrt{2}(5+2a^2)w}
\end{eqnarray}
where $w=\sqrt{\frac{1}{56+9a^2(5+a^2)}}$.
\subsection{Example 3}
Let us consider a two-qutrit isotropic state described by the density operator $\rho_{\beta}$ in (\ref{betastate}).

Let $f_3(x)$ be chracteristic polynomial of $\rho_{\beta}$ given as
\begin{eqnarray}
	f_3(x) &=& x^9 - a_1(\beta) x^8 +a_2(\beta) x^7 - a_3(\beta) x^6 + a_4(\beta) x^5 \nonumber \\&& - a_5(\beta) x^4  + a_6(\beta) x^3 - a_7(\beta) x^2 + a_8(\beta) x - a_{9}(\beta)\nonumber
\end{eqnarray}
where the coefficients $a_i(\beta)$, in terms of moments may be expressed as
\begin{eqnarray}
	&a_1(\beta) = \frac{1}{3}(1+8\beta),  \; \; &a_2(\beta) = \frac{4}{9}f(2+7\beta), \nonumber\\& a_3(\beta) = \frac{28}{27}f^2(1+2\beta),  \; \; &a_4(\beta) = \frac{14}{81}f^3(4+5\beta), \nonumber\\ &a_5(\beta) = \frac{14}{243}f^4(5+4\beta),  \; \; &a_6(\beta) = \frac{28}{729}f^5(2+\beta), \nonumber\\
	&a_7(\beta) = \frac{4\beta^6(7+2\beta)}{2187},  \; \; &a_8(\beta) = \frac{\beta^7(8+\beta)}{6561}
\end{eqnarray}
and $a_9(\beta) =\frac{\beta^8}{19623}$.
Since all the coefficients $a_i(\beta),~~i=1~ \text{to} ~9$ of $f_3(x)$ are positive, realignment matrix $R(\rho_{\beta})$ is positive semi-definite. Thus $\widetilde{R}(\rho_\beta)$ is completely positive for $0 \leq p \leq 1$.
\subsection{Example 4}
Consider the $\alpha$-state defined in (\ref{eg4}). Let $f_2(x)$ be chracteristic polynomial of $\rho_{\alpha}$ given as
\begin{eqnarray}
	f_4(x) &=& x^9 - a_1(\alpha) x^8 +a_2(\alpha) x^7 - a_3(\alpha) x^6 + a_4(\alpha) x^5 \nonumber \\&& - a_5(\alpha) x^4  + a_6(\alpha) x^3 - a_7(\alpha) x^2 + a_8(\alpha) x - a_{9}(\alpha)\nonumber
\end{eqnarray}
where the coefficients $a_i(\alpha)$ calculated in terms of moments using (\ref{rec}) are given as
\begin{eqnarray}
	&a_1(\alpha) = \frac{1 + 17\alpha}{2(1 + 8\alpha)},  \; \; &a_2(\alpha) = \frac{\alpha(7+ 59\alpha)}{2(1+8\alpha)^2}, \nonumber\\& a_3(\alpha) = \frac{a^2(21 + 109\alpha)}{2(1+8\alpha)^3},  \; \; &a_4(\alpha) = \frac{5\alpha^3(7+ 23\alpha)}{2(1+8\alpha)^4}, \nonumber\\ &a_5(\alpha) = \frac{\alpha^4(35 + 67\alpha)}{2(1+8\alpha)^5},  \; \; &a_6(\alpha) = \frac{\alpha^5(21 + 17\alpha)}{2(1+8\alpha)^6}, \nonumber\\
	&a_7(\alpha) = \frac{\alpha^6(7 - \alpha)}{2(1+8\alpha)^7},  \; \; &a_8(\alpha) = \frac{\alpha^7(1-\alpha)}{2(1+8\alpha)^8}
\end{eqnarray}
and $a_9(\alpha) =0$.
Now since $a_i(\alpha) \geq 0$ for $i=1$ to $9$, $R(\rho_{\alpha})$ is PSD. Hence, by $Theorem-1$, $\widetilde{R}(\rho_{\alpha})$ defines a positive map for $0 \leq p \leq 1$ and for all $\alpha \in (0,1)$.	
	 
\section*{Appendix II: Estimation of first moment of $R(\rho)$}
Let $\rho_{AB}$ be a $d\otimes d$ dimensional state. In \cite{sougato}, it has been shown that the measurement of moments of a partial transposed matrix is technically possible using $m$ copies of the state $\rho_{AB}$ and SWAP operations. In this process, the matrix power is written as expectation value of a permutation operator. 
We can apply the same method adopted in the references \cite{horodecki8,ekert} but on the single copy of realigned matrix as 
\begin{eqnarray}
	m_1 &=& Tr[R(\rho_{AB})P] 
\end{eqnarray}
where $P$ is the normalized permutation operator. Now since $R(\rho_{AB})$ is not physically realizable, we need to express the first moment $m_1$ of $R(\rho_{AB})$ in terms of a physically realizable operator. From the definition (\ref{sparealign}) of the SPA of realigned matrix, we can write
\begin{eqnarray}
	R(\rho_{AB}) \propto \widetilde{R}(\rho_{AB}) - \frac{p}{d^2} I_{d \otimes d} 
\end{eqnarray}
Therefore, the first moment of $R(\rho_{AB})$ may be expressed as
\begin{eqnarray}
	m_1 &\simeq& Tr	[(\widetilde{R}(\rho_{AB}) - \frac{p}{d^2} I_{d \otimes d} )P]\nonumber\\
	&=& Tr	[\widetilde{R}(\rho_{AB})P] - \frac{p}{d^2} Tr[P]\nonumber\\
	&=& Tr	[\widetilde{R}(\rho_{AB})P] - \frac{p}{d^2}\nonumber\\
	&\leq& Tr	[\widetilde{R}(\rho_{AB})P] - \frac{ k}{m_1+ d^2k}
	\label{m11}
\end{eqnarray}
In the last line, we have used $p \geq \frac{ d^2k}{m_1+ d^2k} $ and $k=max[0,-\lambda_{min}^{lb}[R(\rho_{AB})]]$, which is defined in Theorem 1. The equality is obtained when all the eigenvalues of $R(\rho_{AB})$ is positive.\\
The inequality (\ref{m11}) may be re-expressed as
\begin{eqnarray}
	m_1 + \frac{ k}{m_1+ d^2k} \leq  Tr	[\widetilde{R}(\rho_{AB})P] := s \label{s1}
\end{eqnarray}
Since $\widetilde{R}(\rho_{AB})$ is a positive semi-definite operator with unit trace, so $s = Tr[\widetilde{R}(\rho_{AB})P]$ can be measured using controlled swap operations [42].\\ 
The inequality (\ref{s1}) can be re-expressed as
\begin{eqnarray}
m_1^2 +m_1(d^2k -s) + k(1-d^2s) \leq 0 \label{quad1}
\end{eqnarray}
Solving the above quadratic equation for $m_1$, we have
\begin{eqnarray}
	\frac{ -(d^2k - s) - \sqrt{(d^2k-s)^2 - 4k(1-d^2s)}}{2}	\leq m_1 \nonumber\\ \leq \frac{ -(d^2k - s) + \sqrt{(d^2k-s)^2 - 4k(1-d^2s)}}{2}
\end{eqnarray}
For $m_{1}$ to be real, we have 
\begin{eqnarray}
	(d^2k-s)^2 - 4k(1-d^2s) \geq 0 
	\label{ineq1}
\end{eqnarray}
Also, and let us assume $1-d^2s \geq 0$. The inequality (\ref{ineq1}) may be further simplified to
\begin{eqnarray}
	(d^2k-s)^2 - 4k(1-d^2s) \geq 0\nonumber\\
	\Rightarrow d^4k^2 + 2k(d^2s -2) + s^2 \geq 0  \label{quad2}
\end{eqnarray}
Inequality (\ref{quad2}) holds when either $k \geq \frac{2 -d^2s+ 2\sqrt{1-d^2s}}{d^4}$ or $k \leq \frac{2 -d^2s - 2\sqrt{1-d^2s}}{d^4}$.\\
\textbf{Case 1:} If $ 2-d^2s+ 2\sqrt{1-d^2s} \leq d^4k \leq d^4 $ then
\begin{eqnarray}
	f_l(s)\leq	m_1 \leq f_u(s) \label{case1}
\end{eqnarray}
\textbf{Case 2:} If $0 \leq d^  4k \leq 2 -d^2s - 2\sqrt{1-d^2s}$ then
\begin{eqnarray}
	g_l(s)\leq	m_1 \leq g_u(s) \label{case2}
\end{eqnarray}
The functions $f_l(s)$, $f_u(s)$, $g_l(s)$, $g_u(s)$ are given as follows:
\begin{eqnarray}
f_l(s) &=&  \frac{1}{2}
(-d^2 + s) \nonumber\\&& -\frac{1}{2d^2} \sqrt{d^8 + 2d^6s + 4d^2s + d^4s^2 -8(1+\sqrt{x})}\nonumber\\
\label{e1}
\end{eqnarray}
\begin{eqnarray}
f_u(s) &=& \frac{-1}{d^2}(x + \sqrt{x}) + \nonumber\\&& \frac{1}{2d^2}\left(\sqrt{d^8 + 2d^6s + 4d^2s +d^4s^2 -8 (1+ \sqrt{x})}\right)\nonumber\\ 
\label{e2}
\end{eqnarray}
\begin{eqnarray}
g_l(s) &=& \frac{1}{d^2}\left(  -x + \sqrt{x} - \sqrt{1+x-2\sqrt{x}} \right) 
\label{e3}
\end{eqnarray}
\begin{eqnarray}
g_u(s) &=& \frac{s}{2} + \frac{1}{d^2} \sqrt{1+x-2\sqrt{x}}
\label{e4}
\end{eqnarray}
where $x = 1 - d^2s$.\\
Hence, the first moment of $R(\rho_{AB})$ may be estimated using (\ref{case1}), (\ref{case2}), (\ref{e1}), (\ref{e2}), (\ref{e3}), (\ref{e4}). Since the functions $f_{l}$, $f_{u}$, $g_{l}$ and $g_{u}$ are expressed in terms of $s= Tr[\widetilde{R}(\rho_{AB})P]$, the first moment of $R(\rho_{AB})$ can be estimated experimentally.

\end{document}